\documentclass[10pt]{article}
\usepackage[dvipdfmx]{graphicx}
\usepackage[normalem]{ulem}
\usepackage{latexsym}
\usepackage{url}
\usepackage{amsmath,mathtools,amssymb}
\usepackage{pifont}
\usepackage{textcomp}
\usepackage[T1]{fontenc}
\usepackage{bbold}

\usepackage{slashbox}
\usepackage{ascmac}

\usepackage{comment}
\usepackage{cancel}

\usepackage{multicol}
\usepackage{multirow}

\usepackage[rm,up,sc,compact,topmarks,calcwidth,pagestyles]{titlesec}

\usepackage[top=2cm,bottom=2cm,left=1cm,right=1.45cm]{geometry}

\def\mbf#1{\mbox{\boldmath ${#1}$}}

\newcommand{\red}[1]{\textcolor{red}{#1}}

\renewcommand\thesection{\Roman{section}}
\titleformat{\section}[hang]{\bfseries\large}{\textbf{\thesection}.}{5pt}{}
\titlespacing{\section}{0pt}{18pt}{6pt}

\titleformat{\subsection}[hang]{\bfseries\large}{\textbf{\Alph{subsection}.}}{5pt}{}
\titlespacing{\paragraph}{0pt}{18pt}{6pt}

\titleformat{\subsubsection}[hang]{\normalsize}{\textbf{\thesubsubsection}}{5pt}{}
\titlespacing{\paragraph}{0pt}{18pt}{6pt}

\titleformat{\paragraph}[hang]{\bfseries\normalsize}{{\theparagraph}}{5pt}{{\ding{110}}\ }
\titlespacing{\paragraph}{0pt}{18pt}{0pt}

\usepackage{subfigure}

\makeatletter
\newcommand{\figcaption}[1]{\def\@captype{figure}\caption{#1}}
\newcommand{\tblcaption}[1]{\def\@captype{table}\caption{#1}}
\makeatother

\usepackage{caption}
\usepackage{fancyhdr}
\providecommand{\url}[1]{\texttt{#1}}

\providecommand{\Capitalize}[1]{\uppercase{#1}}
\providecommand{\capitalize}[1]{\expandafter\Capitalize#1}

\providecommand{\bbletal}{et~al.}

\usepackage[driverfallback=dvipdfm,hyperfootnotes=false]{hyperref}
\hypersetup{
    colorlinks=true,
    citecolor=blue,
    linkcolor=red,
    urlcolor=orange,
}

\begin{document}
\thispagestyle{empty}
\vspace*{8mm}
\begin{center}
  \LARGE{\textbf{Optimization of quantum noise in space gravitational-wave antenna DECIGO with optical-spring quantum locking considering mixture of vacuum fluctuations in homodyne detection\\}}
  \vspace{15mm}
  \large{Kenji Tsuji$^A$, Tomohiro Ishikawa$^A$, Kentaro Komori$^{B,\ C}$, Koji Nagano$^D$, Yutaro Enomoto$^E$,\\ Yuta Michimura$^{B,\ F}$, Kurumi Umemura$^A$, Ryuma Shimizu$^A$, Bin Wu$^A$, Shoki Iwaguchi$^A$,\\Yuki Kawasaki$^A$, Akira Furusawa$^{E,\ G}$, Seiji Kawamura$^{A,\ H}$\\}
  \vspace{8mm}
  \begin{itemize}
    \setlength{\itemsep}{-1pt}
    \item[$^A$] \normalsize{\textit{Department of Physics, Nagoya University, Furo-cho, Chikusa-ku, Nagoya, Aichi 464-8602, Japan\\}}
    \item[$^B$] \normalsize{\textit{Research Center for the Early Universe (RESCEU), School of Science, University of Tokyo, Tokyo 113-0033, Japan\\}}
    \item[$^C$] \normalsize{\textit{Department of Physics, University of Tokyo, Bunkyo, Tokyo 113-0033, Japan\\}}
    \item[$^D$] \normalsize{\textit{LQUOM, Inc., Tokiwadai, Hodogaya, Yokohama city, Kanagawa, 240-8501, Japan\\}}
    \item[$^E$] \normalsize{\textit{Department of Applied Physics, School of Engineering, University of Tokyo, 7-3-1 Hongo, Bunkyo-ku, Tokyo 113-8656, Japan\\}}
    \item[$^F$] \normalsize{\textit{LIGO Laboratory, California Institute of Technology, Pasadena, California 91125, USA\\}}
    \item[$^G$] \normalsize{\textit{Center for Quantum Computing, RIKEN, 2-1 Hirosawa, Wako, Saitama 351-0198, Japan\\}}
    \item[$^H$] \normalsize{\textit{The Kobayashi-Maskawa Institute for the Origin of Particles and the Universe, Nagoya University, Nagoya, Aichi 464-8602, Japan}}
  \end{itemize}
\end{center}
\vspace{5mm}

\begin{abstract}
\normalsize{Quantum locking using optical spring and homodyne detection has been devised to reduce quantum noise that limits the sensitivity of DECIGO, a space-based gravitational wave antenna in the frequency band around 0.1 Hz for detection of primordial gravitational waves. The reduction in the upper limit of energy density ${\Omega}_{\mathrm{GW}}$ from $2{\times}10^{-15}$ to $1{\times}10^{-16}$, as inferred from recent observations, necessitates improved sensitivity in DECIGO to meet its primary science goals. To accurately evaluate the effectiveness of this method, this paper considers a detection mechanism that takes into account the influence of vacuum fluctuations on homodyne detection. In addition, an advanced signal processing method is devised to efficiently utilize signals from each photodetector, and design parameters for this configuration are optimized for the quantum noise. Our results show that this method is effective in reducing quantum noise, despite the detrimental impact of vacuum fluctuations on its sensitivity.}
\normalsize
\end{abstract}

\vspace{15mm}


\section{INTRODUCTION}
\label{Sec:introduction}

Since the first detection of gravitational waves in 2015\ \cite{PhysRevLett.116.061102}, the ongoing observation of gravitational waves resulting from the mergers of binary black holes and binary neutron stars has marked a momentous juncture in the domain of gravitational wave astronomy. This continuous endeavor bestows upon us novel perspectives into astronomical phenomena, catalyzing a revolutionary transformation within the field of astronomy \cite{PhysRevX.11.021053}. Moreover, it is envisaged that the growing significance of gravitational wave detections will play an increasingly pivotal role in the realm of astronomy with contributions from ground-based detectors such as Einstein Telescope \cite{Punturo_2010} and Cosimic Explorer \cite{PhysRevD.91.082001}, or space-based detectors such as LISA \cite{amaroseoane2017laser}. This is owing to their unique ability to discern enigmatic phenomena that pose formidable challenges for observation through electromagnetic waves.

The focus of our research is primordial gravitational waves. If we can directly detect these gravitational waves, which originate from quantum fluctuations in the spacetime during the cosmic inflation, we can gain a more detailed understanding of the early stages of the universe, including confirming the occurrence of cosmic inflation \cite{PhysRevD.90.063513}. Detectors exhibiting high sensitivity in the frequency band around 0.1 Hz, which are free of seismic noise from the Earth and thermal noise from mirror suspension, are required. However, conventional ground-based detectors, LIGO \cite{Aasi_2015}, VIRGO \cite{Acernese_2015} and KAGRA \cite{Somiya_2012} cannot remove them.

Our project for these detections is called the DECi-hertz Interferometer Gravitational-wave Observatory (DECIGO) and has been promoted as a space-based gravitational wave antenna in Japan\ \cite{PhysRevLett.87.221103,10.1093/ptep/ptab019}. DECIGO stands out due to its distinctive utilization of three drag-free satellites deployed in space and a Fabry-Perot interferometer configuration, spanning a length of 1000 km, to mitigate the adverse effects of earth's seismic noise and thermal noise from mirror suspensions. This configuration therefore allows us to target a lower frequency band (in this case the 0.1 Hz band) than the frequency band where ground-based detectors have high sensitivity, typically around 100 Hz. Meanwhile, recent analysis of observations by the Planck satellite and others\ \cite{akrami2020planck} have shown that the original design of DECIGO was not sensitive enough to detect primordial gravitational waves. Hence, developing techniques to improve the sensitivity of DECIGO is required.
Previous studies have proposed optimizing the design parameters of DECIGO \cite{galaxies9010009, galaxies9010014, galaxies10010025}, as well as a technique known as quantum locking \cite{PhysRevLett.90.083601,Antoine Heidmann_2004}, to enhance sensitivity by mitigating quantum noise. Quantum noise predominantly affects the low-frequency band and arises from the quantum fluctuations of laser light. Notably, quantum locking involves incorporating sub-cavities on both sides that share a mirror with the 1,000 km primary cavity, resulting in a reduction of radiation pressure noise. Given the high optical loss and lack of squeezing feasibility in the main cavity of DECIGO, this method proves to be exceptionally effective \cite{YAMADA2020126626}, and its utility has been progressively confirmed through in-principle verification experiments \cite{PhysRevD.107.022007}. Additionally, the configuration termed optical-spring quantum locking, which employs optical springs \cite{1970JETP...31..829B} and homodyne detection in sub-cavities for quantum locking, has theoretically exhibited a considerable enhancement in sensitivity \cite{YAMADA2021127365}.

However, prior investigations have not employed homodyne detection based on an experimental setup that faithfully reflects the actual optical effects, rather assumed ideal homodyne detection. This limitation arises due to the treatment of parameter $\eta$, responsible for determining the direction of homodyne detection, which has not been ascertained through interference light but rather regarded as an independent variable subject to arbitrary choice during simulations. In order to faithfully model the system based on the actual setup, it becomes imperative to account for the influence of the mixture of vacuum fluctuations. Consequently, the determination of the homodyne detection direction necessitates the extraction of light from the conventional optical path to produce interference, wherein the use of beam splitters, acting as points of interference, is indispensable. Hence, the primary objective of this paper is the accurate evaluation of the homodyne detection approach, taking into consideration the mixture of vacuum fluctuations. Furthermore, in this configuration, an additional photodetector can be employed, and we propose a method to utilize the signal from this supplementary photodetector. In this paper, we apply this method while optimizing each design parameter in a manner akin to previous research \cite{YAMADA2021127365}.

In Section \ref{Sec:optical design}, a more elaborate elucidation of the optical design is presented, encompassing the mitigation of quantum noise through optical-spring quantum locking, along with the configuration of homodyne detection incorporating these constituents. The intricate approach to signal optimization, achieved through the combination of multiple signals, is expounded upon in Section \ref{sec:Method of Signal Processing}. Furthermore, this section offers a comprehensive block diagram illustrating the acquisition of these signals. Sections \ref{SEc:simulations} through \ref{Sec:result and discussion} showcase simulation results, demonstrating the parameters that yield the utmost sensitivity for DECIGO under this configuration, as well as the corresponding obtained signals. Additionally, we delve into the sensitivity difference of DECIGO in comparison with prior research.

\begin{figure}[t]
  \centering
  \includegraphics[width=150mm]{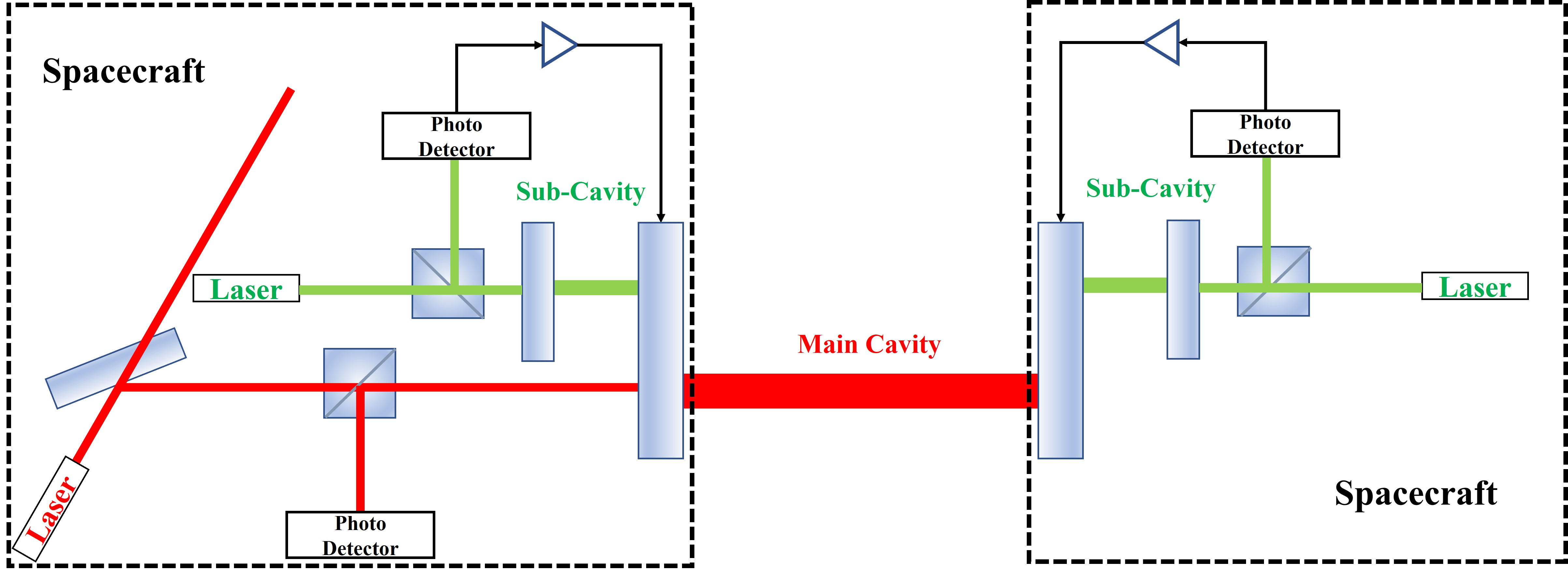}
  \caption{Standard configuration of quantum locking in DECIGO. The path that the red laser light passes through represents the main cavity, while the paths that the green laser lights pass through correspond to the sub-cavities. The two mirrors of the main cavity are shared with the sub-cavities, and the photodetector signals from the sub-cavities contain information regarding the noise induced by the main laser light.}
  \label{fig:Quantum Locking}
\end{figure}
\section{OPTICAL DESIGN }\label{Sec:optical design}

\subsection{Optical-Spring Quantum Locking}
In this subsection, we describe optical-spring quantum locking used to reduce quantum noise.
Initially, we elucidate the methodology employed to address quantum fluctuations, utilizing a mathematical framework known as quadrature-phase amplitude \cite{schleich2011quantum}. This formalism incorporates the application of creation and annihilation operators, denoted as $a_{j}$ and $a_j^{\dag}$, respectively, which satisfy the commutation relation (Eq. \ref{eq:commutator}). These operators play a pivotal role in the analysis of the optical mode within each cavity.
\begin{equation}\label{eq:commutator}
  [a_{j},a_{j^{\prime}}^{\dag}] = 2{\pi}{\delta}_{jj^{\prime}}
\end{equation}
Here, $j$ is an identifier of the three cavities, indicating Main(=m), Sub1(=s1) or Sub2(=s2). Then, using $a_{j}$ and $a_j^{\dag}$ operators, the amplitude quantum fluctuation $q_j$ and phase quantum fluctuation $p_j$ are defined by
\begin{equation}\label{eq:quantum_fluctuations}
  q_{j} = {\frac{1}{\sqrt{2}}}(a_j{}+a_j^{\dag})\hspace{2mm},\hspace{2mm}p_{j} = {\frac{1}{i\sqrt{2}}}(a_j-a_j^{\dag}).
\end{equation}
We now elaborate on the approach of reducing quantum noise through quantum locking. Figure \ref{fig:Quantum Locking} illustrates the standard configuration of quantum locking designed for DECIGO. In this configuration, supplementary sub-cavities, equipped with shorter cavity lengths and shared mirrors, are appended to both sides of the main optical cavity, which spans a distance of 1,000 km between DECIGO satellites. Notably, the laser sources employed for the sub-cavities differ from the one utilized for the main cavity. Based on this setup, crucial information concerning the amplitude quantum fluctuations of the main laser light can be extracted from the signals transmitted via the sub-cavity laser light. By regulating the shared mirror using these signals, it becomes possible to effectively nullify the radiation pressure noise stemming from the main cavity, while simultaneously preserving the underlying gravitational wave signal.

Moreover, we  utilize the optical spring for the sub-cavities. The optical spring is a contrived technology that accomplishes heightened sensitivity by creating the interaction between mechanical and optical effects. A mirror positioned slightly away from the resonance point experiences a constant external force that counterbalances the radiation force within the cavity. Consequently, when the radiation force changes due to quantum fluctuations, the mirror behaves akin to one mounted on a spring due to its relationship with the radiation force.

\begin{figure}[t]
  \centering
  \includegraphics[width=140mm]{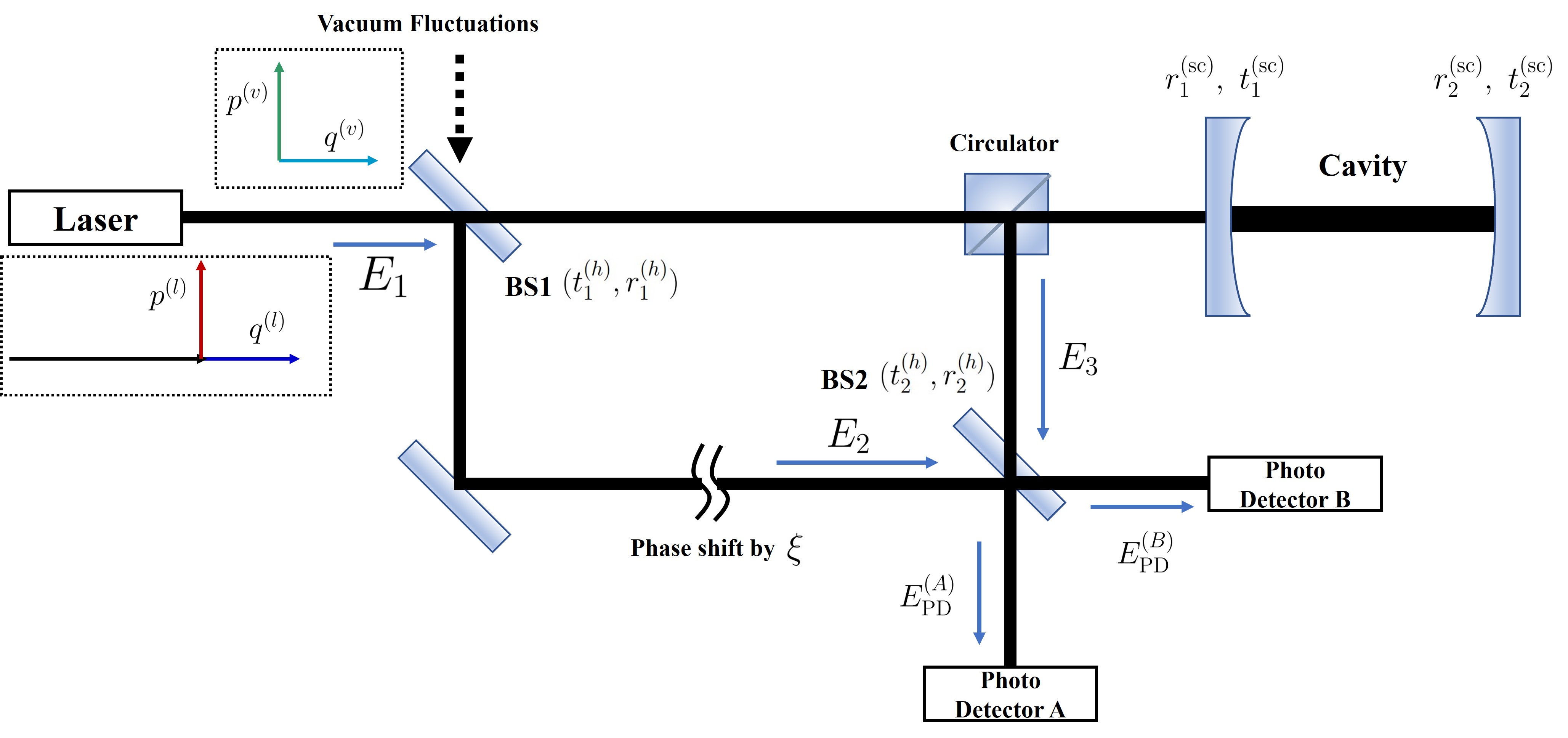}
  \caption{Concept of homodyne detection. The lower path represents the optical path for the local oscillator, which undergoes a phase shift by $\xi$. $E_i$ is the classical electric field of the laser light. The regions enclosed by dotted lines are quantum fluctuations of the laser right or vacuum fluctuations. We use superscript $(l)$ to denote quantum fluctuations of the laser light and superscript $(v)$ to denote vacuum fluctuations, distinguishing their respective noise sources.}
  \label{fig:hd_path}
\end{figure}

\subsection{Homodyne Detection}
In addition to employing optical-spring quantum locking, we adopt homodyne detection to mitigate the influence of quantum noise. The concept of homodyne detection is elucidated in Figure \ref{fig:hd_path}. In this configuration, beam splitter 1 (=BS1) is strategically positioned to extract light as a local oscillator from the incident light directed towards the cavity. Meanwhile, beam splitter 2 (=BS2) brings the local oscillator light to interference with the light that either reflects or traverses the cavity. This arrangement allows for detection along a direction different from that of the normal carrier light, playing a crucial role in quantum noise reduction. The interfered light is subsequently captured by two photodetectors, as depicted in the figure. The variables ${\eta}^{(A)}$ and ${\eta}^{(B)}$, which determine the detection axis angle, are derived through the ensuing steps. To commence, let us delineate the electric field $E_1$ of the light emanating from the laser source as follows:
\begin{equation}\label{eq:E1}
  E_1 = E_0e^{i{\omega}_0t},
\end{equation}
where $E_0$ is a constant representing the amplitude of the electric field, and ${\omega}_0$ is the angular frequency of the laser light. At the same time, the local oscillator $E_2$ is obtained using the amplitude reflectance $r^{(h)}_1$ of BS1, given by
\begin{equation}
  E_2 = r^{(h)}_1e^{i{\xi}}E_0e^{i{\omega}_0t} = r^{(h)}_1e^{i{\xi}}E_1.
\end{equation}
Here, $\xi$ is a parameter defined as the relative phase shift associated with the change in the optical path length. The light directed towards the cavity combines the light entering the cavity with the light reflected by the input mirror, and the carrier light $E_3$ is expressed as follows:
\begin{equation}
  E_3  = \Bigl[-r_1^{(sc)}+{\frac{{t_1^{(sc)}}^2r_2^{(sc)}e^{i{\phi}^{\prime}}}{1-r_1^{(sc)}r_2^{(sc)}e^{i{\phi}^{\prime}}}}\Bigr]t_1^{(h)}E_1.
\end{equation}
Here,  $r^{(h)}_1$ represents the amplitude transmittance of BS1, while $r_1^{(sc)},t_1^{(sc)}$ and $r_2^{(sc)},t_2^{(sc)}$ correspond to the amplitude reflectance or the amplitude transmittance of the mirrors used in the sub-cavities. For the subsequent calculations, we set $r_2^{(sc)}$ to 1, and $t_2^{(sc)}$ to 0. Besides, ${\phi}^{\prime}$ is defined as the change in laser phase when the laser light completes one round trip through the cavities. Consequently, the field falling on the photodiodes $E_{\mathrm{PD}}^{(A)},E_{\mathrm{PD}}^{(B)}$ by lights passing through each optical path is given as follows:
\begin{equation}
  \begin{split}
    E_{PD}^{(A)} &= t_2^{(h)}E_3+r_2^{(h)}E_2\\
    &=\Bigl[t_1^{(h)}t_2^{(h)}\Bigl\{-r_1^{(sc)}+{\frac{{t_1^{(sc)}}^2r_2^{(sc)}}{(1-r_1^{(sc)}r_2^{(sc)})^2+4r_1^{(sc)}r_2^{(sc)}{\sin^2{\frac{{\phi}^{\prime}}{2}}}}}(e^{i{\phi}^{\prime}}-r_1^{(sc)})\Bigr\}+r_1^{(h)}r_2^{(h)}e^{i{\xi}}\Bigr]E_1\\
    &= \Biggl[\Bigl\{t_1^{(h)}t_2^{(h)}\Bigl(-r_1^{(sc)}+{\frac{{t_1^{(sc)}}^2r_2^{(sc)}(\cos{{\phi}^{\prime}}-r_1^{(sc)})}{(1-r_1^{(sc)}r_2^{(sc)})^2+4r_1^{(sc)}r_2^{(sc)}{\sin^2{\frac{{\phi}^{\prime}}{2}}}}}\Bigr)+r_1^{(h)}r_2^{(h)}{\cos{\xi}}\Bigr\}\\
    &\hspace{6cm}+i\Bigl\{t_1^{(h)}t_2^{(h)}{\frac{{t_1^{(sc)}}^2r_2^{(sc)}{\sin{{\phi}^{\prime}}}}{(1-r_1^{(sc)}r_2^{(sc)})^2+4r_1^{(sc)}r_2^{(sc)}{\sin^2{\frac{{\phi}^{\prime}}{2}}}}}+r_1^{(h)}r_2^{(h)}{\sin{\xi}}\Bigr\}\Biggl]E_1\\
    &\equiv (A_1+iA_2)E_1
  \end{split}
\end{equation}
\begin{equation}
  \begin{split}
    E_{PD}^{(B)} &= -r_2^{(h)}E_3+t_2^{(h)}E_2\\
    &= \Bigl[-t_1^{(h)}r_2^{(h)}\Bigl\{-r_1^{(sc)}+{\frac{{t_1^{(sc)}}^2r_2^{(sc)}}{(1-r_1^{(sc)}r_2^{(sc)})^2+4r_1^{(sc)}r_2^{(sc)}{\sin^2{\frac{{\phi}^{\prime}}{2}}}}}(e^{i{\phi}^{\prime}}-r_1^{(sc)})\Bigr\}+r_1^{(h)}t_2^{(h)}e^{i{\xi}}\Bigr]E_1\\
    &=\Biggl[\Bigl\{-t_1^{(h)}r_2^{(h)}\Bigl(-r_1^{(sc)}+{\frac{{t_1^{(sc)}}^2r_2^{(sc)}(\cos{\phi}^{\prime}-r_1^{(sc)})}{(1-r_1^{(sc)}r_2^{(sc)})^2+4r_1^{(sc)}r_2^{(sc)}{\sin^2{\frac{{\phi}^{\prime}}{2}}}}}\Bigr)+r_1^{(h)}t_2^{(h)}{\cos{\xi}}\Bigr\}\\
    &\hspace{6cm}+i\Bigl\{-t_1^{(h)}r_2^{(h)}{\frac{{t_1^{(sc)}}^2r_2^{(sc)}{\sin{{\phi}^{\prime}}}}{(1-r_1^{(sc)}r_2^{(sc)})^2+4r_1^{(sc)}r_2^{(sc)}{\sin^2{\frac{{\phi}^{\prime}}{2}}}}}+r_1^{(h)}t_2^{(h)}{\sin{\xi}}\Bigr\}\Biggr]E_1\\
    &\equiv (B_1+iB_2)E_1.\\
  \end{split}
\end{equation}
Hence, through the consideration of the electric field $E_1$ pertaining to the laser emission at its immediate inception as the point of reference, each instance of interference light is detected along an axis rotated at an angle ascertained by the subsequent equation:

\begin{align}\label{eq:eta}
  {\eta}^{(A)} = {\arctan{\frac{A_1}{A_2}}},\hspace{5mm}{\eta}^{(B)} = {\arctan{\frac{B_1}{B_2}}}.
\end{align}

Next, we consider the quantum state at this juncture. If the quantum state of the laser light manifests as a coherent state, then the light extracted as a local oscillator, influenced solely by the amplitude reflectance $r_1^{(h)}$ of BS1, also assumes a coherent state. On the contrary, the light emanating from the cavity to the BS2 exhibits a squeezed state. This arises due to the amplitude fluctuations of the laser light, which result in mirror displacements and changes in the optical path length, thereby causing amplitude fluctuations to manifest as phase fluctuations. This phenomenon is referred to as ponderomotive squeezing \cite{PhysRevD.23.1693}. Consequently, the interference light resulting from the combination of these two beams is squeezed in its quantum state. Thus, quantum noise can be ameliorated by judiciously selecting an appropriate angle for the detection axis, as defined in Eq. \ref{eq:eta}. The optimal angle is determined such that the projective component of the quantum fluctuation along this axis nullifies each other.

Incidentally, it should be noted that this particular configuration embraces vacuum fluctuations at BS1. Omitting consideration of vacuum fluctuations would lead to a reduction in the quantum fluctuation of the light transmitted or reflected by BS1. Hence, we introduce a symmetrical position with respect to the beam splitter, as illustrated in the figure, wherein vacuum fluctuations are injected. This serves to compensate for the reduction in quantum fluctuation of the laser light.
\section{Signal Processing}\label{sec:Method of Signal Processing}

\subsection{Block Diagram}
\begin{figure}[t]
  \centering
  \includegraphics[width=160mm]{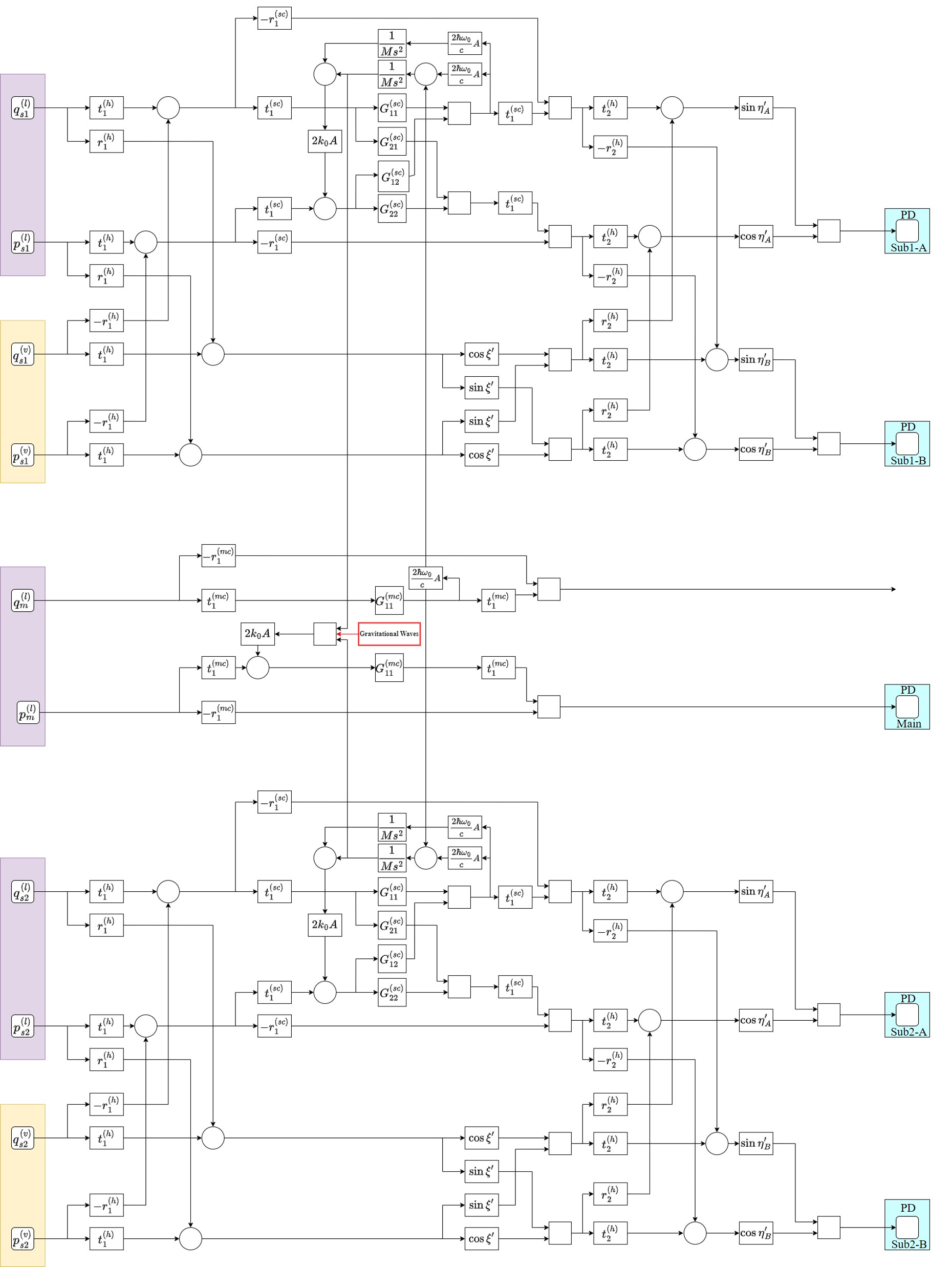}
  \caption{Block Diagram of the detection system incorporating optical-spring quantum locking and accounting for the effects of vacuum fluctuations. The upper and lower sections are the sub-cavities, and the central section is the main cavity. Purple areas represent quantum fluctuations of the laser light, yellow areas represent vacuum fluctuations mixed in during homodyne detection, and cyan areas represent detection ports.}
  \label{fig:Block Diagram}
\end{figure}

\hspace{5mm} We utilize a block diagram to portray the interferometer configuration shown in Fig. \ref{fig:Quantum Locking}, which integrates optical-spring quantum locking and accounts for the amalgamation of vacuum fluctuations in homodyne detection, as depicted in Fig. \ref{fig:Block Diagram}. This schematic comprises three distinct sections: the upper and lower segments represent the sub-cavities, while the central segment denotes the main cavity. The purple and yellow regions on the left side correspond to quantum fluctuations of the laser light and the mixed vacuum fluctuations in homodyne detection, respectively. Within these regions, the upper port signifies amplitude quantum fluctuations, whereas the lower port pertains to phase fluctuations. On the right side, the cyan region designates the detection port.

The amplitude transmittance and amplitude reflectance of a mirror are defined using the identifier $(k)$, such as (sc) representing the sub-cavity, and the identifier $n$ representing 1 or 2, as follows
\begin{equation}
  \bigl[{r_n^{(k)}}\bigr]^2+\bigl[{t_n^{(k)}}\bigr]^2 = 1.
\end{equation}
Note that this equation assumes no loss effects for all mirrors. In addition, all mirrors used in each cavity have the same mass. In the diagram, the mirrors of the main cavity, which are shared with the sub-cavities, are depicted as blocks located in the upper or lower parts of the sub-cavities. In addition, $G_{ij}$ denotes the matrix of optical-spring effects, which is determined by the detuning angle $\phi$ and sideband frequency $\Omega$, and can be expressed as:
\begin{align}
  g(\Omega) &= \Bigl[1-r_{1}r_{2}e^{-i({\phi}+{\frac{2L}{c}}{\Omega})}\Bigr]^{-1}\\\label{eq:G-matrix}
  G(\Omega) &= {\frac{1}{2}}
  \begin{bmatrix}
     g^{\ast}(\Omega)+g(-{\Omega}) & i[g^{\ast}(\Omega)-g(-{\Omega})] \\
     -i[g^{\ast}(\Omega)-g(-{\Omega}) ] & g^{\ast}(\Omega)+g(-{\Omega})
  \end{bmatrix}
  ,
\end{align}
where $c$ is the speed of light taken as $3{\times}10^8{\mathrm{\, m/s}}$, and $L$ is the cavity length. Next, ${\eta}^{\prime}_{A/B}$ represents the angle between the phase direction of transmitted or reflected light by BS2 in $E_3$ and the axis determined by Eq. \ref{eq:eta}, and it is defined as follows
\begin{equation}
  {\eta}^{\prime}_{A/B} = {\frac{\pi}{2}}+{\eta}^{(A/B)}-{\eta}_0^{(A/B)},
\end{equation}
where ${\eta}_0^{(A/B)}$ denotes the angular orientation of each carrier light with respect to the phase direction of $E_1$. The influence of gravitational waves is introduced into the system within the region enclosed by the red line positioned at the diagram's center. This region corresponds to the displacement of the shared mirror induced by gravitational waves. It is worth noting that the impact of gravitational waves on the sub-cavities is not taken into consideration, given their relatively abbreviated cavity lengths, resulting in negligible alterations in the optical path length caused by gravitational waves. Hence, gravitational
\clearpage
\clearpage
\noindent waves are detected in the phase direction by the photodetector situated at the central portion of the diagram, which captures the main laser light.

Likewise, by utilizing this block diagram, we can ascertain the amplification or attenuation of individual quantum fluctuations and their subsequent detection as noise. Table \ref{table:each signal} showcases the transfer functions of this system, elucidating the correlation between mirror displacement caused by gravitational waves or each quantum fluctuation and the signals obtained by the five photodetectors. Within this table, the gravitational wave signal is denoted by $x$, while $p$ and $q$, as defined in the preceding subsection, represent the bases of the quantum fluctuations. By employing these bases and transfer functions, $V_{\mathrm{main}}$, the signal acquired from the photodetector for the main laser light, can be expressed as follows:
\begin{equation}
  V_{\mathrm{main}} = a\mbf{x}+Aq_m^{(l)}+Bp_m^{(l)}+Cq_{s1}^{(l)}+Dp_{s1}^{(l)}+Cq_{s2}^{(l)}+Dp_{s2}^{(l)}+Eq_{1}^{(v)}+Fp_{1}^{(v)}+Eq_{2}^{(v)}+Fp_{2}^{(v)}.
\end{equation}
In the same way, $V_{\mathrm{sub1}}^{(A)},V_{\mathrm{sub1}}^{(B)},V_{\mathrm{sub2}}^{(A)},V_{\mathrm{sub2}}^{(B)}$ which are the signals obtained from the other four photodetectors, are defined by the combinations shown in the table. Here, note that these signals contain some common noise components, as the sub-cavities on both sides are assumed to have the same configuration.

\begin{table}[t]
  \centering
  \caption{Transfer functions from the effects of gravitational waves or quantum fluctuations to each photodetector.}
  \label{table:each signal}
  \begin{tabular}{cl|ccccccccccc}
    \hline
    \multicolumn{2}{l|}{\backslashbox[28mm]{To}{From}} & $\mbf{x}$ (gw) & $q_m^{(l)}$ & $p_{m}^{(l)}$ & $q_{s1}^{(l)}$ & $p_{s1}^{(l)}$ & $q_{s2}^{(l)}$ & $p_{s2}^{(l)}$ & $q_1^{(v)}$ & $p_1^{(v)}$ & $q_2^{(v)}$ & $p_2^{(v)}$\\
    \hline
    \hline
                             & Main   & $a$ & $A$ & $B$ & $C$ & $D$ & $C$ & $D$ & $E$ & $F$ & $E$ & $F$ \\
    \cline{2-13}
    Photo                    & Sub1-A & $-$ & $G$ & $H$ & $I$ & $J$ & $-$ & $-$ & $K$ & $L$ & $-$ & $-$\\
    \multirow{2}{*}{Detector}& Sub2-A & $-$ & $G$ & $H$ & $-$ & $-$ & $I$ & $J$ & $-$ & $-$ & $K$ & $L$\\
                             & Sub1-B & $-$ & $M$ & $N$ & $O$ & $P$ & $-$ & $-$ & $Q$ & $R$ & $-$ & $-$\\
                             & Sub2-B & $-$ & $M$ & $N$ & $-$ & $-$ & $O$ & $P$ & $-$ & $-$ & $Q$ & $R$\\
    \hline
  \end{tabular}
\end{table}


\subsection{Completing the Square}\label{subsec:Completing the Square}
\hspace{5mm}Completing the square in quantum locking is a signal optimization method that has shown high effectiveness in previous research for reducing quantum noise\ \cite{YAMADA2020126626}. In this approach, a new signal $V = V_1+{\chi}V_2$ is defined from the two signals $V_1$ and $V_2$, with the combination coefficient $\chi$ chosen to minimize the power spectrum. To effectively utilize the signals from homodyne detection at two ports, we introduce additional combination coefficients and devise a method to optimize the combination of three signals. Thus, by using $V_1$ ,$V_{2}^{(A)}$ and $V_{2}^{(B)}$ along with the combination coefficients ${\chi}_A$ and ${\chi}_B$, a new signal $V$ is defined as follows:
\begin{equation}
  V = V_1 + {\chi}_{A}V_{2}^{(A)}+{\chi}_{B}V_{2}^{(B)}.
\end{equation}
The power spectrum $S_V$ of this signal given by:
\begin{equation}\label{eq:opt signal2}
  \begin{split}
    S_V &= \left|V_1\right|^2+\left|V_{2}^{(A)}\right|^2\left|{\chi}_A+{\frac{V_{2}^{{(A)\dag}}V_1+{\chi}_BV_{2}^{(A)\dag}V_{2}^{(B)}}{V_{2}^{(A)\dag}V_{2}^{(A)}}}\right|^2\\
    &\hspace{5mm}+\Biggl(\left|V_{2}^{(B)}\right|^2-{\frac{\left|V_{2}^{(A)}V_2^{(B)}\right|^2}{\left|V_{2}^{(A)}\right|^2}}\Biggr)\left|{\chi}_B+{\frac{\left|V_{2}^{(A)}\right|^2\Bigl(V_{2}^{(B)\dag}V_1\Bigr)-\Bigl(V_{2}^{(A)}V_{2}^{(B)\dag}\Bigr)\Bigl(V_1V_{2}^{(A)\dag}\Bigr)}{\left|V_{2}^{(A)}\right|^2\left|V_{2}^{(B)}\right|^2-\left|V_{2}^{(A)\dag}V_2^{(B)}\right|^2}}\right|^2\\
    &\hspace{5mm}-{\frac{\left|V_1^{\dag}V_{2}^{(A)}\right|^2\left|V_{2}^{(B)}\right|^2+\left|V_1^{\dag}V_{2}^{(B)}\right|^2\left|V_{2}^{(A)}\right|^2-\Bigl(V_{2}^{(B)\dag}V_1\Bigr)\Bigl(V_{2}^{(A)\dag}V_{2}^{(B)}\Bigr)\Bigl(V_1^{\dag}V_{2}^{(A)}\Bigr)-\Bigl(V_{2}^{(B)}V_1^{\dag}\Bigr)\Bigl(V_{2}^{(A)}V_{2}^{(B)\dag}\Bigr)\Bigl(V_1V_{2}^{(A)\dag}\Bigr)}{\left|V_{2}^{(A)}\right|^2\left|V_{2}^{(B)}\right|^2-\left|V_{2}^{(A)\dag}V_2^{(B)}\right|^2}}.
  \end{split}
\end{equation}
Therefore, the values of ${\chi}_A^{\mathrm{(opt)}}$ and ${\chi}_B^{\mathrm{(opt)}}$ that minimize its power spectrum are determined as follows:
\begin{align}
  {\chi}_A^{\mathrm{(opt)}} &= -{\frac{\left|V_{2}^{(B)}\right|^2\Bigl(V_{2}^{(A)\dag}V_1\Bigr)-\Bigl(V_{2}^{(B)}V_{2}^{(A)\dag}\Bigr)\Bigl(V_1V_{2}^{(B)\dag}\Bigr)}{\left|V_{2}^{(A)}\right|^2\left|V_{2}^{(B)}\right|^2-\left|V_{2}^{(A)\dag}V_2^{(B)}\right|^2}}\\
  {\chi}_B^{\mathrm{(opt)}} &= -{\frac{\left|V_{2}^{(A)}\right|^2\Bigl(V_{2}^{(B)\dag}V_1\Bigr)-\Bigl(V_{2}^{(A)}V_{2}^{(B)\dag}\Bigr)\Bigl(V_1V_{2}^{(A)\dag}\Bigr)}{\left|V_{2}^{(A)}\right|^2\left|V_{2}^{(B)}\right|^2-\left|V_{2}^{(A)\dag}V_2^{(B)}\right|^2}}.
\end{align}
Finally, we utilize the three signals as follows:
\begin{equation}
  V_1 = V_{\mathrm{main}}/a\hspace{1mm},\hspace{5mm}V_2^{(A)} = \Bigl[V_{\mathrm{sub1}}^{(A)}+V_{\mathrm{sub2}}^{(A)}\Bigr]/a\hspace{1mm},\hspace{5mm}V_2^{(B)} = \Bigl[V_{\mathrm{sub1}}^{(B)}+V_{\mathrm{sub2}}^{(B)}\Bigr]/a,
\end{equation}
where $a$ is a transfer function from the displacement of the shared mirror caused by gravitational waves to the main photodetector as shown in Tab. \ref{table:each signal}. Each signal is divided by $a$ to calibrate it with the gravitational wave signal.


\section{SIMULATIONS}\label{SEc:simulations}

In this section, we delineate the parameter conditions employed in the simulation, as well as the method utilized for evaluating the sensitivity of DECIGO. Table \ref{table:parameter condition} exhibits the symbols and ranges/values employed in the simulation. The upper segment of this table presents the five variable parameters employed for optimizing sensitivity: $t_1^{(h)}$ and $t_2^{(h)}$ represent the amplitude transmittance of each beam splitter. Detuning angle reflects the effect of the optical spring, and finesse corresponds to the effective number of light reflections within the cavity. Additionally, $\xi$ is a parameter introduced to modulate the length of the optical path taken by the local oscillator. Since the two photodetectors are symmetrically positioned, the A and B signals can be interchanged by swapping the amplitude transmittance and reflectance values of BS2 and adjusting the parameter $\xi$ to become $\xi + \pi$. Note that the range of $\xi$ is therefore defined as between 0 and $\pi$. The eight parameters presented in the lower part of the table remain constant, with the exception of the cavity length, which differs between the main cavity and the sub-cavities.
\begin{table}[t]
  \centering
  \caption{Conditions of main parameters for the optimization of the DECIGO's sensitivity.}
  \label{table:parameter condition}
  \begin{tabular}{lll}
    \hline
    \hline
    Meaning & Symbol  &  \multicolumn{1}{c}{Range/Value}\\
    \hline
    \hline
    \multirow{2}{*}{Amplitude Transmittance} & $t_1^{(h)}$ & $0$ to $1$\\
    &$t_2^{(h)}$ & $0$ to $1$\\[2mm]
    Detuning Angle & ${\phi}_{\mathrm{sub}}$ & $-{\pi}$ to $\pi$\ rad\\[2mm]
    Finesse & ${\mathcal{F}}_{\mathrm{sub}}$ & $1$ to $100$\\[2mm]
    Phase Shift & ${\xi}$  & $0$ to $\pi$\ rad\\[2mm]
    \hline
    \multirow{2}{*}{Cavity Length} & $L_{\mathrm{sub}}$ & $1$\ m\\
    &$L_{\mathrm{main}}$ & $1000$\ km\\[2mm]
    \multirow{2}{*}{Laser Power} & $I_{\mathrm{sub}}$ & $100$\ W\\
    &$I_{\mathrm{main}}$ & $100$\ W\\[2mm]
    \multirow{2}{*}{Laser Wavelength} & ${\lambda}_{\mathrm{sub}}$ & $515$\ nm\\
    &${\lambda}_{\mathrm{main}}$ & $515$\ nm\\[2mm]
    \multirow{2}{*}{Mirror Mass} & $M_{\mathrm{sub}}$ & $100$\ kg\\
    &$M_{\mathrm{main}}$ & $100$\ kg (shared)\\[2mm]
    \hline
    \hline
  \end{tabular}
  \vspace{5mm}
  \caption{Parameters used to calculate the SNR \cite{YAMADA2021127365}.}
  \label{table:Parameters used to calculate the SNR}
  \begin{tabular}{lll}
    \hline
    Meaning & Symbol  &  \multicolumn{1}{c}{Value}\\
    \hline
    Hubble Parameter & $H_0$ & $70{\mathrm{\,km{\cdot}sec^{-1}{\cdot}Mpc^{-1}}}$\\
    Time for Correlation & T & 3 years\\
    Frequency & $f$ & 0.1 to 1 Hz\\
    Correlation Function & ${\gamma}$ & 1\\
    Energy Density & ${\Omega}_{\mathrm{GW}}$ & $10^{-16}$\\
    Noise Power Spectral Densities &$P_1,P_2$ &\\
    \hline
  \end{tabular}
\end{table}
Next, we adopt Signal-to-Noise Ratio (SNR) as a measure to evaluate the sensitivity of DECIGO to primordial gravitational waves. The SNR is given by\ \cite{PhysRevD.59.102001}:
\begin{equation}\label{eq:SNR}
  \mathrm{SNR} = {\frac{3H_0^2}{10{\pi}^2}}{\sqrt{T}}\Bigl[{\int_{0.1}^1}df{\frac{2{\gamma}(f)^2{\Omega}^2_{\mathrm{GW}}(f)}{f^6P_1(f)P_2(f)}}\Bigr]^{\frac{1}{2}},
\end{equation}
where $P_1$ and $P_2$ represent spectral densities of noise, as computed following the methodology outlined in Section \ref{sec:Method of Signal Processing}. Notably, in this case, $P_1$ and $P_2$ assume equal values. The remaining parameters employed in the SNR computation are presented in Table \ref{table:Parameters used to calculate the SNR}. In Eq. \ref{eq:SNR}, $T$ signifies the observation period, which has been set to a duration of 3 years. ${\Omega}_{\mathrm{GW}}$ denotes the energy density of primordial gravitational waves, and for the purpose of this research, a fixed value of $10^{-16}$ has been adopted. Furthermore, $\gamma$ denotes the overlap reduction function \cite{PhysRevD.59.102001}, assumed to be unity in the configuration of DECIGO. Additionally, the SNR assessment specifically concentrates on the frequency band spanning from $0.1$ Hz to $1$ Hz, wherein DECIGO is expected to exhibit high sensitivity to gravitational waves.

\section{RESULT AND DISCUSSION}\label{Sec:result and discussion}
\begin{figure}[t]
  \begin{minipage}[c]{0.65\textwidth}
    \centering
    \includegraphics[width=120mm]{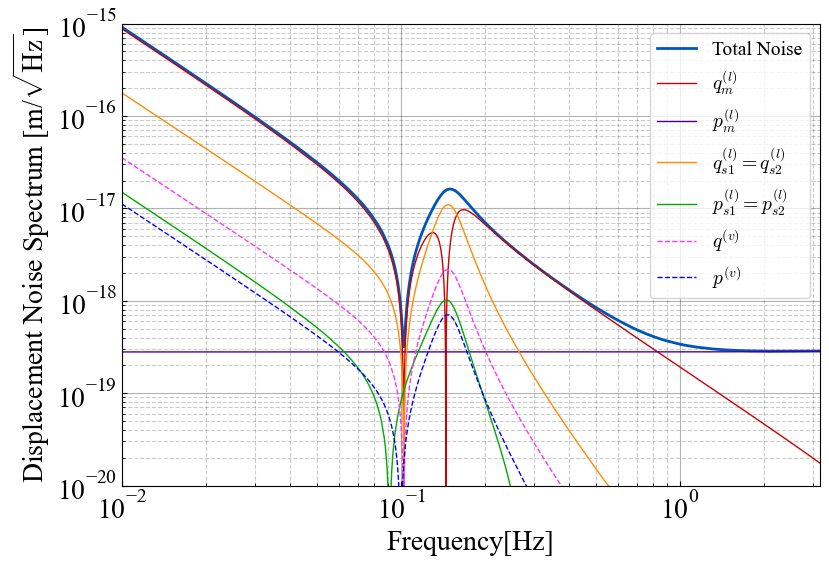}
    \caption{Optimized sensitivity curves with contributions from noise sources. The total noise is represented by the thick blue line, while each quantum noise caused by fluctuations of the laser light or vacuum fluctuations is shown as thin lines. Especially, quantum noises originating from vacuum fluctuations are depicted as dashed lines. Additionally, the green or yellow lines represent signals obtained from only sub-cavity 1. Therefore, note that when adding the signal from sub-cavity 2 to this figure, the curve will be scaled up by a factor of $\sqrt{2}$.}
    \label{fig:opt1}
  \end{minipage}
  \hfill
  \def\@captype{table}
  \begin{minipage}[b]{0.33\textwidth}
    \begin{center}
      \tblcaption{Optimized parameters.}
      \label{table:optimized parameters}
      \begin{tabular}{ll}
        \hline
        \multicolumn{1}{c}{Symbol}  &  \multicolumn{1}{c}{Value}\\
        \hline
        $t_1^{(h)}$ & $0.152$\\[2mm]
        $t_2^{(h)}$ & $0.308$\\[2mm]
        ${\phi}_{\mathrm{sub}}$ & $5.90{\times}10^{-3}$\ rad\\[2mm]
        ${\mathcal{F}}_{\mathrm{sub}}$ & $55.3$\\[2mm]
        ${\xi}$  & $0.260$\ rad\\[2mm]
        \hline
        &\\
        \multicolumn{1}{c}{SNR} & \multicolumn{1}{c}{$79.6$}\\[3mm]
        \hline
      \end{tabular}
    \end{center}
  \end{minipage}
\end{figure}
\begin{figure}[t]
  \vspace{-5mm}
  \centering
  \includegraphics[width=120mm]{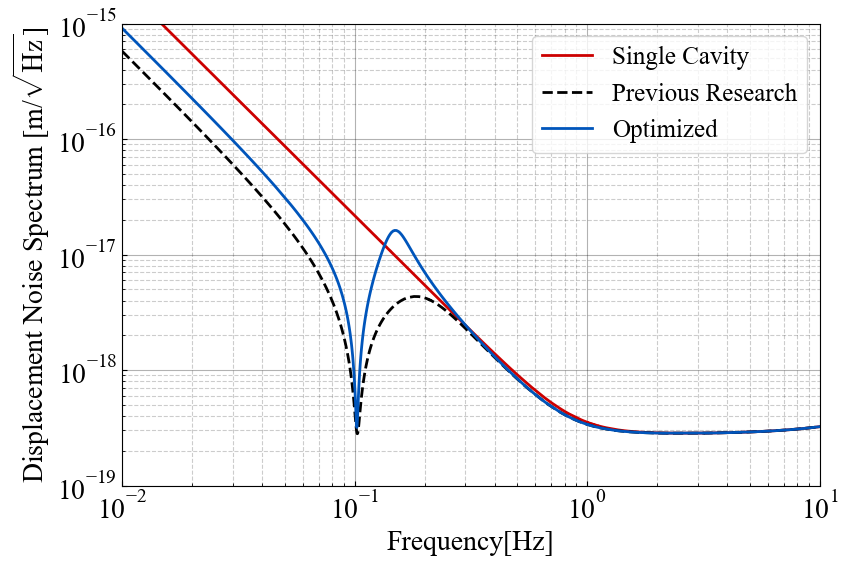}
  \caption{Optimized sensitivity curves considering noise contributions. The blue line represents the sensitivity curve optimized in this study, which corresponds to the total noise in Fig. \ref{fig:opt1}. The dashed line represents the sensitivity curve optimized in the previous research \cite{YAMADA2021127365}, resulting in an SNR of 141. The red line represents the optimized sensitivity curve when using a single cavity with a length of 1000 km.}
  \label{fig:cr1}
\end{figure}

Based on the simulation results, we obtained the optimized sensitivity curves of DECIGO with optical-spring quantum locking to detect primordial gravitational waves as shown in Fig. \ref{fig:opt1}. Each sensitivity curve, excluding the shot noise caused by phase quantum fluctuations of the main laser light, exhibits a dip around 0.1 Hz, and these dips align with each other. On the other hand, the shot noise, which has no dip, remains nearly constant within the target frequency band and contributes to the baseline of DECIGO's sensitivity. Furthermore, Table \ref{table:optimized parameters} shows the parameters that minimize noise in relation to the gravitational wave signal, resulting in a calculated SNR of 79.6. The more refined model presented in this paper gives a different result than less-detailed modeling (SNR=141), as shown in Fig. \ref{fig:cr1}. \footnote{Here, note that this differs from the results (SNR=214) described in the previous paper \cite{YAMADA2021127365}. The previous paper used approximations; a constant amplitude of the light is assumed even in the case of off-resonant condition. In this paper, we calculated the SNR without any approximations, and obtained SNR of 141. } In this paper, the local oscillator is extracted from the incoming light directed towards the cavity. As a result, the laser light going to the sub-cavities are smaller, and the noises are relatively higher due to the effects entering from the vacuum fluctuation. In addition, the local light must have a large value in order to obtain an arbitrary homodyne angle. In the previous research, we considered a hypothetical situation in which the local laser power is infinite, independent of the incoming light directed towards the cavity. However, since it is not possible to do so in reality, we obtained degraded but realistic sensitivity. In contrast, the SNR without optical-spring quantum locking is 1.74, indicating that optical-spring quantum locking significantly helps reduce quantum noise even when the effect of vacuum fluctuations is considered.

\section{CONCLUSIONS}
To assess the efficacy of homodyne detection in sub-cavities within the optical-spring quantum locking for mitigating noise in DECIGO, it was imperative to consider a detector configuration that accounts for the vacuum fluctuations. In this paper, we investigated the determination of the homodyne angle by employing extracted light serving as a local oscillator, and subsequently constructed a comprehensive block diagram incorporating vacuum fluctuations. Moreover, this particular configuration enables the utilization of two photodetectors, which sets it apart from previous DECIGO configurations that allowed only the use of a single photodetector. We presented an optimization method for processing the signals acquired from each photodetector to obtain the optimized sensitivity curve. By optimizing the parameters with these configurations, the impact of vacuum fluctuations was evaluated. Although the effect of vacuum fluctuation results in a sensitivity slightly inferior to that of the ideal situation shown in the previous research, it still exhibits a marked improvement compared to the scenario without optical-spring quantum locking. Consequently, we demonstrate that homodyne detection in optical-spring quantum locking is an effective technique for noise reduction in DECIGO, with the potential to significantly contribute to the observation of primordial gravitational waves.\\[5mm]


\noindent
\large
\textbf{Acknowledgements}\\
\normalsize
We would like to thank David H. Shoemaker for commenting on a draft. This work was supported by JSPS KAKENHI, Grants No. JP19H01924 and No. JP22H01247. This work was also supported by Murata Science Foundation.







\end{document}